\documentclass[preprint,10pt,number]{elsarticle}
\usepackage[a4paper,left=25mm,right=25mm,top=25mm,bottom=25mm]{geometry}
\usepackage{setspace}
\usepackage{amssymb}
\usepackage{amsmath}
\usepackage{amsfonts}
\usepackage{graphicx} 
\usepackage{hyperref}
\usepackage{enumerate}

\usepackage{amsthm}
\theoremstyle{remark}

\usepackage{xcolor}
\usepackage{ulem}

\hypersetup{colorlinks=true, allcolors=blue}

\usepackage{autonum}
\definecolor{deeppink}{HTML}{ff1493}
\definecolor{royalblue}{HTML}{4169e1}
\definecolor{gainsboro}{HTML}{dcdcdc}

\newcommand{\rev}[1]{{\color{black}#1}}

\journal{Physica A}

\begin{document}

\begin{frontmatter}



\title{Scale-free congestion clusters in large-scale traffic networks: a continuum modeling study}

\author[utmath]{Yuki Chiba}
\author[utmath]{Norikazu Saito}
\author[utmath]{Yuki Ueda\fnref{fn1}}
\author[tytlabs]{Hiroaki Yoshida\fnref{fn2}}

\address[utmath]{Graduate School of Mathematical Sciences, The University of Tokyo,\\ 3-8-1 Komaba, Meguro-ku, Tokyo, 153-8914, Japan}
\address[tytlabs]{Toyota Central R\&D Labs., Inc., Nagakute, Aichi, 480-1192, Japan}
\fntext[fn1]{Present address: Research Institute for Electronic Science, Hokkaido University, Sapporo 060-0812, Japan}
\fntext[fn2]{Correspondence to: h-yoshida@mosk.tytlabs.co.jp}

\begin{abstract}
Recent empirical studies have reported that spatiotemporal congestion clusters in urban traffic exhibit scale-free statistics, with cluster size following a power-law distribution. In this study, we address whether macroscopic continuum descriptions of traffic flow are capable of generating such scale-free spatiotemporal congestion patterns. To this end, we analyze the second-order Aw-Rascle-Zhang model on directed networks under junction coupling.
The governing equations are solved by a high-order discontinuous Galerkin scheme, and junction fluxes are determined by an optimization-based coupling procedure enforcing conservation and admissibility at intersections. 
Congestion is defined by thresholding the road-averaged density, and spatiotemporal clusters are extracted as connected components in space and time.
Numerical experiments on lattice networks of varying sizes reveal that the cluster size follows a robust power-law distribution.
Moreover, when rescaled by the linear system size inherent to the two-dimensional network geometry, the distribution collapses onto an approximately universal curve, indicating finite-size scaling governed by the linear system size.
The observed power-law statistics and finite-size scaling are reminiscent of scale-invariant dynamics characteristic of self-organized criticality.
These results demonstrate that macroscopic continuum traffic models can reproduce large-scale statistical features observed in real urban congestion dynamics.
\end{abstract}




\end{frontmatter}



\section{Introduction}
Urban traffic congestion has intensified worldwide and now exceeds pre-pandemic levels in many major cities. According to recent reports~\cite{inrix2024}, drivers lose dozens of hours annually to congestion, with substantial economic and environmental costs. Congestion is therefore recognized as a complex urban phenomenon affecting mobility, emissions, and economic activity. Correspondingly, research on traffic dynamics has expanded across multiple modeling scales. Comprehensive reviews such as Helbing’s survey of self-driven many-particle systems~\cite{helbing2001} and Kerner’s three-phase traffic theory~\cite{kerner2004} summarize decades of microscopic and macroscopic investigations. Empirical studies have further demonstrated large-scale traffic patterns, including the macroscopic fundamental diagram observed in Yokohama~\cite{geroliminis2008} and network-wide spatiotemporal analyses based on probe-vehicle data~\cite{saberi2020,Zhao2025}. Together, these works highlight the growing emphasis on understanding congestion as a large-scale collective phenomenon.

While the concept of resilience, i.e., quantifying the persistence of congestion, has attracted attention across various fields \cite{holling1973, chang2004, gao2016}, Zhang~\textit{et~al.}~\cite{zhang2019} analyzed link-speed data from Beijing and Shenzhen and reported that both the cluster size and the recovery time follow power-law distributions. 
This observation suggests that urban congestion exhibits features reminiscent of self-organized criticality (SOC)~\cite{bak1988,Chen2024}. Similar scaling behavior had previously been discussed in traffic models~\cite{Linesch2008,nagatani1995}. These findings naturally raise the broader question of whether such scale-free statistics can be reproduced within theoretical models of traffic flow.

\begin{figure}[tb]
    \centering 
    \includegraphics[clip, scale=0.8]{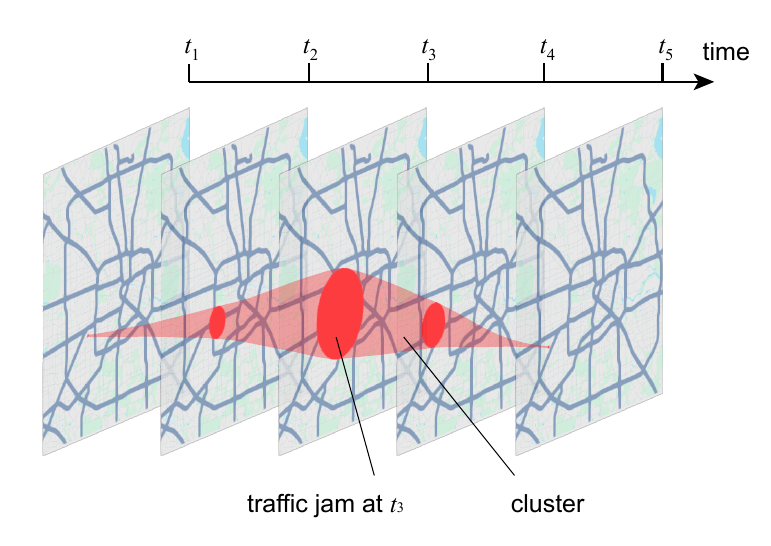} 
    \caption{Schematic illustration of congestion clusters. Congested regions on the network (red) at times $t_1,\dots,t_5$ are shown, indicating that a traffic jam at a given time (e.g., $t_3$) forms a cluster as a set connected in space and time.}    \label{Fig:intro}
\end{figure}

Congestion mechanisms have been extensively studied using discrete microscopic formulations, particularly cellular automata, as well as through microsimulation platforms such as \textsc{SUMO}~\cite{krajzewicz2012sumo,sumo2023}, which resolve individual vehicle dynamics in detail. 
While these approaches capture fine-scale behavior, their vehicle-level resolution and modeling complexity can hinder systematic investigations of network-scale statistical structures.
In contrast, macroscopic traffic dynamics are commonly described by hyperbolic conservation laws, following the foundational works of Lighthill-Whitham and Richards~\cite{LighthillWhitham1955,Richards1956}.
\rev{
Some continuum studies have addressed network-scale properties and large-scale traffic dynamics \cite{Marigo2009, Mollier2019}, and the broader theory of continuum flows on networks, including traffic applications, has been reviewed by Bressan et al.~\cite{Bressan2014}.}
The Aw-Rascle-Zhang (ARZ) model~\cite{AwRascle2000,ZHANG2002} provides a widely used second-order continuum formulation incorporating driver anticipation. 
Because continuum models offer a coarse-grained yet dynamically consistent representation of traffic at the network scale, they provide a natural framework for investigating the emergence of scale-free congestion statistics.
When continuum traffic models are extended to urban networks, a central technical challenge lies in prescribing consistent coupling conditions at junctions where multiple roads merge and diverge. 
\rev{ For first-order traffic flow models on networks, Canic et al. developed a Runge–Kutta DG framework with junction coupling conditions~\cite{Canic2015}. For second-order traffic flow, Buli and Xing proposed a DG method for the Aw–Rascle model on networks~\cite{BuliXing2020}. The present study adopts the latter continuum-modeling viewpoint, in which both density and velocity are evolved, and uses it to investigate spatiotemporal congestion-cluster statistics on a large directed network.}

\rev{
In the present study, we investigate the emergence and scaling structure of spatiotemporal congestion clusters within a coarse-grained continuum network formulation. Using the Aw-Rascle-Zhang model combined with optimization-based junction coupling, we first compute the time-dependent density field on directed networks using a high-order discontinuous Galerkin scheme. The obtained network-scale density evolution is used to identify congested links by thresholding the road-averaged density and to construct spatiotemporal congestion clusters as connected components in space and time. We then analyze the statistical properties of these clusters, with particular emphasis on their size distribution and finite-size scaling behavior across networks of varying sizes.
}

While previous empirical and discrete modeling studies have reported scale-free congestion statistics, it remains unclear whether such behavior requires microscopic stochasticity or agent-based interactions. In this study, we show that even a macroscopic continuum model, when posed on a complex network with physically consistent junction coupling, can generate scale-free spatiotemporal congestion clusters with finite-size scaling structure. This provides evidence that scale invariance may arise as an intrinsic property of nonlinear networked traffic dynamics at the continuum level.

This paper is organized as follows.
In Section~\ref{sec:model}, we introduce the continuum traffic flow model, the network configuration, and the coupling conditions imposed at junctions. Section~\ref{sec:simulation} describes the DG discretization and the numerical framework. In Section~\ref{sec:results}, we present numerical results on spatiotemporal congestion clusters, analyze the resulting power-law statistics, and examine finite-size scaling behavior across networks of varying sizes.
Finally, in Section~\ref{sec:clusterstats}, we summarize the paper and outline directions for future research. 
Some technically more involved material is deferred to the appendices. In \ref{sec:math}, we
provide supplementary remarks on the continuum mathematical model from the viewpoint of
hyperbolic systems, while \ref{sec:coupling} presents a precise formulation of the coupling conditions.

\section{The continuum model}
\label{sec:model}

\rev{
In this section, we formulate traffic flow on a directed road network within a continuum framework. We introduce the Aw-Rascle-Zhang model on each road, together with the network configuration and the coupling and boundary conditions at junctions and boundaries. This formulation provides the mathematical basis for the numerical scheme in the next section and for the congestion-cluster analysis presented later.
}

\subsection{The ARZ traffic flow model}

We first consider an infinitely long single-lane road on which
infinitely many vehicles are traveling in one direction.
Let $\rho(x,t)>0$ denote the vehicle density and $v(x,t)>0$ the velocity,
both assumed to be smooth functions. The ARZ traffic flow model is given by
\begin{equation}
   \label{Eq:AR_model}
    \left\{\begin{array}{rl}
        \partial_t \rho + \partial_x(\rho v) &= 0, \\[2mm]
        \partial_t\bigl(\rho(v+p)\bigr) + \partial_x\bigl(\rho (v+p)v\bigr) &= 0,
    \end{array}\right.
\end{equation}
where $p$ denotes the ``pressure'' of the traffic flow and is interpreted as an anticipation factor
that determines the velocity from the density. Henceforth, we assume that
$p=p(\rho)$ is a given smooth, increasing function of $\rho$.

The first equation in Eq.~\eqref{Eq:AR_model} expresses the so-called mass conservation law, which
states that the total number of vehicles is conserved. In the classical continuum traffic flow
model known as the Lighthill--Whitham--Richards  model \cite{LighthillWhitham1955, Richards1956},
only this first equation is considered, and the velocity $v$ is assumed to be a prescribed function
of the density $\rho$. This model is commonly referred to as a first-order model. As an improvement of these classical models, systems have been proposed in
which one does not assume an {a priori} function relationship between $v$ and $\rho$.
Instead, an additional equation is introduced to determine $v$, leading to a coupled system. These models are referred to as second-order models. Among such models, a well-known example is the Payne--Whitham model
\cite{Payne1971,Whitham1974}, in which the second equation is derived by imposing a momentum
conservation law, by analogy with one-dimensional fluid models. However, it is a trivial but important fact that traffic flow,
unlike a fluid, does not satisfy a momentum conservation law. To address this issue, Aw and Rascle \cite{AwRascle2000} proposed setting the material derivative of $v+p$ equal to zero:
\begin{equation}
    \label{Eq:material_derivative_is_zero}
    \frac{D}{Dt}(v+p)
    = \partial_t (v+p) + v\,\partial_x (v+p) = 0.
\end{equation}
By rewriting this equation using the first equation of the system, one obtains the second equation
in Eq.~\eqref{Eq:AR_model}. This second equation represents a conservation law for
\[
    q = \rho (v+p).
\]
As pointed out in \cite{AwRascle2000}, this quantity does not possess any particular physical meaning.
On the other hand, \cite{BuliXing2020} refers to $q$ as ``pseudo-momentum''. 

Almost simultaneously with the work of \cite{AwRascle2000}, Zhang \cite{ZHANG2002} formalized fundamental principles of traffic flow, namely, that drivers react only to conditions ahead and that the speed of information propagation does not exceed the vehicle speed, and derived the same system of PDEs. Consequently, the model \eqref{Eq:AR_model} is now commonly referred to as the ARZ model.

Since the velocity is given by
\begin{equation}
\label{eq:velocity}
    v = \frac{q}{\rho} - p,
\end{equation}
the ARZ model~\eqref{Eq:AR_model} can be equivalently written as
\begin{equation}
\label{Eq:AR_model2}
    \left\{\begin{array}{rl}
        \partial_t \rho + \partial_x\bigl(q - \rho p(\rho)\bigr) &= 0, \\[2mm]
        \partial_t q + \partial_x\left(\dfrac{q^2}{\rho} - p(\rho)\,q\right) &= 0.
    \end{array}\right.
\end{equation}

For later convenience, we represent this system in terms of vectors.
Let
\begin{equation}
    \label{eq:conservation_of_laws0}
    U = \begin{pmatrix}
        \rho \\
        q
    \end{pmatrix}, \qquad
    F(z) = \begin{pmatrix}
        z_2 - z_1 p(z_1) \\
        \dfrac{z_2^2}{z_1} - z_2 p(z_1)
    \end{pmatrix}
    \quad \text{for} \quad
    z = \begin{pmatrix}
        z_1 \\
        z_2
    \end{pmatrix}.
\end{equation}
Then Eq.~\eqref{Eq:AR_model2} can be rewritten as
\begin{equation}
    \label{eq:conservation_of_laws}
    \partial_t U + \partial_x F(U) = 0.
\end{equation}


\subsection{Networks}

\begin{figure}[tb]
    \centering 
    \includegraphics{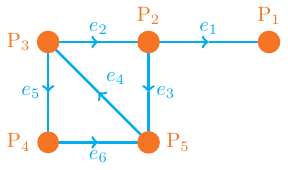} 
    \caption{
    An example of the network geometry considered in this paper. $\mathcal{J}=\{\mathrm{P}_2,\mathrm{P}_3,\mathrm{P}_4,\mathrm{P}_5\}$,  
    $\mathcal{B}=\{\mathrm{P}_1\}$, 
    $\mathrm{Out}_3=\{e_2,e_5\}$, and $\mathrm{In}_3=\{e_4\}$.  
    }
    \label{Fig:network}
\end{figure}

In this paper, we consider traffic flow modeled by the ARZ model \eqref{Eq:AR_model} on a complex network rather than on a single road. The network considered in this paper is not an abstract graph in the sense of graph theory, but a geometric object consisting of points and line segments embedded in $\mathbb{R}^2$. A typical example is displayed in Fig.~\ref{Fig:network}. 
More precisely, it is defined as follows.

Let $\mathcal{N}=\{\mathrm{P}_1,\mathrm{P}_2,\ldots,\mathrm{P}_{\nu}\}$ be a finite set of distinct points in $\mathbb{R}^2$, which we call \emph{nodes}. 
Let $\mathcal{E}=\{e_1,e_2,\ldots,e_{\mu}\}$ be a finite collection of line segments in $\mathbb{R}^2$, which we call \emph{roads} (or edges). 
Each road $e_k$ connects two nodes $\mathrm{P}_j$ and $\mathrm{P}_l$ and is endowed with an orientation, that is, $\mathrm{P}_j$ is the source (initial node), and $\mathrm{P}_l$ is the target (terminal node) of $e_k$.

For a node $\mathrm{P}_j$, we define
\[
\mathrm{In}_j = \{ e_k \in \mathcal{E} \mid \mathrm{P}_j \text{ is the target of } e_k \},\qquad 
\mathrm{Out}_j = \{ e_k \in \mathcal{E} \mid \mathrm{P}_j \text{ is the source of } e_k \}.
\]
Clearly, $\mathrm{In}_j$ and $\mathrm{Out}_j$ are subsets of $\mathcal{E}$.

We further assume that any two distinct roads $e_k$ and $e_m$ either do not intersect or share exactly one common node. 
A node $\mathrm{P}_j$ satisfying $|\mathrm{In}_j|>0$ and $|\mathrm{Out}_j|>0$ is called a \emph{junction}, where $|\mathrm{In}_j|$ and $|\mathrm{Out}_j|$ denote the numbers of roads belonging to $\mathrm{In}_j$ and $\mathrm{Out}_j$, respectively. 
In what follows, we call $\mathrm{P}_j$ an $N_I$--$N_O$ junction if 
$N_I=|\mathrm{In}_j|$ and $N_O=|\mathrm{Out}_j|$. 
A node $\mathrm{P}_l$ satisfying either $(|\mathrm{In}_l|=0,\ |\mathrm{Out}_l|=1)$ or $(|\mathrm{In}_l|=1,\ |\mathrm{Out}_l|=0)$ is called a \emph{boundary node}. 
We denote by $\mathcal{J}$ and $\mathcal{B}$ the sets of all junctions and boundary nodes, respectively, and assume that
\[
\mathcal{N} = \mathcal{J} \cup \mathcal{B}.
\]

Under this setting, each road $e_k$ is identified with the interval $[0,l_k]$ on the real-line $\mathbb{R}$, where $l_k$ denotes the length $|e_k|$ of $e_k$. 
The coordinate $x=0$ corresponds to the source of $e_k$, while $x=|e_k|$ corresponds to its target. 
On each road, we solve the ARZ model \eqref{Eq:AR_model}. 

\subsection{Coupling and boundary conditions}

To describe traffic flow on the entire network, it is necessary to prescribe coupling conditions at junctions and boundary conditions at boundary nodes. 
The boundary conditions are relatively simple. 
Indeed, we impose time-periodic Dirichlet boundary conditions at the boundary nodes. 
This allows us to control the inflow of vehicles into the network and the outflow to the exterior, thereby inducing a nonstationary evolution of the total number of vehicles within the network.

On the other hand, the coupling conditions at junctions are more delicate. 
We essentially follow the approach proposed in the previous work~\cite{BuliXing2020}. 
For the reader's convenience, we briefly review it below.

The numerical scheme in \cite{BuliXing2020} is based on the mathematical analysis for the unique existence of the solution to Riemann problem at junctions. 
First, it is quite natural to assume the conservation of mass (number of vehicles) at junctions, that is: 
\begin{itemize} 
    \item [(C1)] The flux of the density must be conserved. 
\end{itemize}
Furthermore, the following assumption is often accepted as a reasonable rule. 
\begin{itemize} 
    \item [(C2)] The waves produced at the junction must have negative speed in incoming roads and positive speed in outgoing road. 
\end{itemize}

In general, these coupling conditions cannot provide a unique solution to the Riemann problem for an $N_I$--$N_O$ junction. 
We need additional conditions to the flux of the density at the junction:

\begin{itemize} 
    \item [(C3)] The density on each incoming road allocated to outgoing roads according to given ``traffic distribution matrix'' $A \in\mathbb{R}^{N_I\times N_O}$.
    \item [(C4)] The sum of flux of density in incoming roads at the junction must be maximized. 
\end{itemize}

The condition (C4) can provide a unique solution on incoming roads. 
On the other hand, we need additional rule to determine the unique solution at outgoing roads. 
One of the proposed choices in \cite{BuliXing2020} is:
\begin{itemize}
    \item[(C5)] Maximize the velocity at each outgoing road. 
\end{itemize}

The descriptions of (C4) and (C5) given above are somewhat abstract. In \ref{sec:coupling}, we explain
their concrete meaning using explicit mathematical expressions.

In \cite{BuliXing2020}, the authors proposed two numerical schemes: one based on the coupling conditions introduced above, and the other based on the conditions proposed in \cite{HautBastin2007}. They reported that the numerical results obtained with the two schemes are nearly identical, except in cases where a capacity drop occurs. 
\rev{ 
In this paper, we adopt the scheme proposed in \cite{BuliXing2020}, which applies the coupling conditions (C1)--(C5).

Since the choice of well-posed coupling conditions for the ARZ model on networks is not unique, the present results should be understood as those obtained under one representative and physically consistent junction formulation. At the same time, the comparison reported in \cite{BuliXing2020} suggests that, as mentioned above, the qualitative features of the numerical results are not strongly sensitive to the specific admissible coupling choice. Our purpose here is therefore not to perform a systematic comparison of alternative junction rules, but to examine whether scale-free congestion statistics can emerge in a continuum network model under such a formulation.
}


\section{Model discretization and computational algorithm}
\label{sec:simulation}

In this section, we present the discretization of the ARZ model \eqref{Eq:AR_model} on each edge under the coupling conditions~(C1)--(C5), and the computational algorithm. 
The discretization and the handling of the coupling conditions are basically based on the method proposed in~\cite{BuliXing2020}. 
However, to enhance numerical stability and facilitate large-scale network simulations, 
we modify the computational approach proposed in~\cite{BuliXing2020} in several aspects.
First, the coordinate direction on each road is chosen to coincide with the flow direction rather than the junction-based orientation.
Second, we simplify the treatment of the constraints.
In~\cite{BuliXing2020}, the constraints for all roads at a junction are combined, leading to up to six coupled equations.
In contrast, we treat the constraints for each road separately.
As a result, the resulting minimization problem involves solving at most two equations per road. We explain the procedure in detail below.

In what follows, we assume the following relation between the pressure and the density:
\begin{gather}
p(\rho) = \rho^{\gamma}\quad \mbox{ for some }\gamma> 0,
\label{eq:p1}\\
\rho_{\mathrm{max}} = 1,\quad v_{\mathrm{max}} = p(\rho_{\mathrm{max}}) = \rho_{\mathrm{max}}^{\gamma} = 1,
\label{eq:p2}
\end{gather}
where $\rho_{\mathrm{max}} > 0$ and $v_{\mathrm{max}} > 0$, respectively, denote the
 prescribed maximal density and the prescribed maximum velocity. 

\subsection{Spatial discretization}

Let $e_k\in \mathcal{E}$ and consider the ARZ model \eqref{Eq:AR_model} on $(0,l_k)$, where $l_k=|e_k|$. Under assumption \eqref{eq:p1}, the ARZ model is expressed as Eq.~\eqref{eq:conservation_of_laws} with 
\[
F(U)= \begin{pmatrix}
    q-\rho^{\gamma+1}\\
     \frac{q^2}{\rho}-q\rho^\gamma
\end{pmatrix}
.
\]

We introduce a subdivision of the interval $(0,l_k)$ into small subintervals
\[
I_j=(x_{j-1/2},x_{j+1/2}), \qquad j=1,\ldots,N_k,
\]
and define the finite element space $V_{h,k}$ by
\[
V_{h,k}
=
\bigl\{
U_h \in L^2((0,l_k))^2
\mid 
\mbox{For each $I_j$, each component of $U_h$ is affine on $I_j$}
\bigr\}.
\]

For $U_h,W_h\in V_{h,k}$, we write 
\[
(U_h,W_h)_{I_j}=\int_{I_j}U_h(x)W_h(x)~dx.
\]

We further define the one-sided traces at the cell interfaces by
\[
U_{h,j+1/2}^- := \lim_{x\to x_{j+1/2}-0} (U_h|_{I_j})(x),
\qquad
U_{h,j+1/2}^+ := \lim_{x\to x_{j+1/2}+0} (U_h|_{I_{j+1}})(x).
\]

For $U_h \in V_{h,k}$ and given boundary values
$U_{h,1/2}^-,\, U_{h,N_k+1/2}^+ \in \mathbb{R}^2$,
we define $\tilde L(U_h) \in V_{h,k}$ by the following DG scheme:
find $\tilde L(U_h) \in V_{h,k}$ such that
\begin{multline}
    (\tilde L(U_h),\phi_h)_{I_j}
    =
    (F(U_h),\partial_x \phi_h)_{I_j}
    - \hat F_{j+1/2}\bigl(U_{h,j+1/2}^-,U_{h,j+1/2}^+\bigr)\,\phi_{h,j+1/2}^- \\
    + \hat F_{j-1/2}\bigl(U_{h,j-1/2}^-,U_{h,j-1/2}^+\bigr)\,\phi_{h,j-1/2}^+,
    \qquad
    j=1,\ldots,N_k,\ \phi_h \in V_{h,k},
    \label{eq:dg_scheme}
\end{multline}
where the numerical flux $\hat F_{j+1/2}$ is given by the Lax--Friedrichs flux
\[
\hat F_{j+1/2}(U^-,U^+)
=
\frac12\bigl(F(U^-)+F(U^+)\bigr)
-\frac{\alpha}{2}\,(U^+ - U^-),
\]
and $\alpha>0$ is the Lax--Friedrichs constant.

Finally, in order to reduce numerical instabilities, we apply the TVB limiter; see~\cite{Shu2009} for details. 
We denote by $L(U_h) \in V_{h,k}$ the result obtained after applying the TVB limiter to $\tilde L(U_h)$.

\subsection{Time discretization}

Let $\Delta t>0$ be a given time step. 
Given $U_h^i \in V_{h,k}$, we compute the solution at the next time level 
$U_h^{i+1} \in V_{h,k}$ using the third-order strong stability preserving Runge--Kutta method (SSPRK3); see, for example,~\cite{GottliebShuTadmor2001}. 
Specifically, we compute
\begin{align}
    U_{h}^{(1)} &= U_h^i + \Delta t\, L(U_h^i), \label{eq:ssprk1} \\
    U_{h}^{(2)} &= \frac{3}{4}U_h^i + \frac{1}{4}U_h^{(1)} 
    + \frac{1}{4}\Delta t\, L(U_h^{(1)}), \label{eq:ssprk2} \\
    U_{h}^{i+1} &= \frac{1}{3}U_h^i + \frac{2}{3}U_h^{(2)} 
    + \frac{2}{3}\Delta t\, L(U_h^{(2)}). \label{eq:ssprk3}
\end{align}

\subsection{Boundary conditions for the DG scheme on a road}

In order to implement the DG scheme presented in the previous subsections on a road $e_k$, 
we need to specify the boundary values $U_{h,1/2}^-$ and $U_{h,N_k+1/2}^+$. 
If the source (respectively, the target) of $e_k$ is a boundary node, 
we impose Dirichlet boundary conditions at $x=0$ (respectively, $x=l_k$). 
At junctions, the boundary values are determined by introducing ghost-cell states whose values are computed from the coupling conditions.

\begin{figure}[tb]
    \centering
    \includegraphics[clip, scale=1.2]{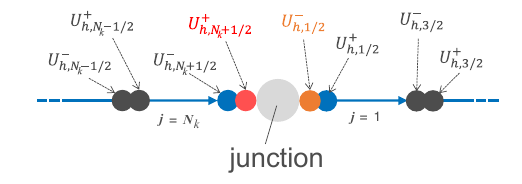}
    \caption{Schematic illustration of the junction boundary treatment in the DG scheme. 
    At the cell interfaces $j=1/2$ and $j=N_k+1/2$, 
    the left and right traces of the numerical solution $U_{h,j\pm 1/2}^{\pm}$ 
    and the ghost-side states determined from the coupling conditions (shown in red) are distinguished.}
    \label{Fig:boundary}
\end{figure}

A schematic illustration of the junction treatment in the DG scheme is shown in Fig.~\ref{Fig:boundary}. 
At each junction (e.g., at $j=1/2$ and $j=N_k+1/2$), 
the left and right traces of the numerical solution, denoted by $U_{h,j\pm1/2}^{\pm}$, are distinguished. 
The ghost-side states (shown in red in Fig.~\ref{Fig:boundary}$)$ are computed by solving a local optimization problem that enforces the coupling conditions together with admissible wave directions.

At a junction $\mathrm{P}_j\in\mathcal{J}$, if $e_k \in \mathrm{In}_j$, we define
\[
D_{\textup{in}}^{(k)}(U_{h,N_k+1/2}^+)
=
-\hat F_{N_k+1/2}(U_{h,N_k+1/2}^-,U_{h,N_k+1/2}^+)
+ F(U_{h,N_k+1/2}^+),
\]
whereas if $e_k \in \mathrm{Out}_j$, we set
\[
D_{\textup{in}}^{(k)}(U_{h,1/2}^-)
=
\hat F_{1/2}(U_{h,1/2}^-,U_{h,1/2}^+)
- F(U_{h,1/2}^-).
\]

We determine the ghost-state values by minimizing 
$\|D_{\textup{in}}^{(k)}\|_{\mathbb{R}^2}$ 
subject to the constraint $\Phi^{(k)}(U_I^{(k)})=0$. 
Here, for $e_k \in \mathrm{In}_j$, we define
\begin{gather}
U_I^{(k)}
=
\begin{pmatrix}
\rho_I^{(k)}\\
q_I^{(k)}
\end{pmatrix}
=
\frac12\bigl(U_{h,N_k+1/2}^- + U_{h,N_k+1/2}^+\bigr)
+ \frac{1}{2\alpha}\bigl(F(U_{h,N_k+1/2}^+)-F(U_{h,N_k+1/2}^-)\bigr), \\
\Phi^{(k)}(U_I^{(k)}) 
=
q_I^{(k)} - (\rho_I^{(k)})^{\gamma+1} - \hat\delta_{j,k},
\end{gather}
while for $e_k \in \mathrm{Out}_j$, we set
\begin{gather}
U_I^{(k)}
=
\begin{pmatrix}
\rho_I^{(k)}\\
q_I^{(k)}
\end{pmatrix}
=
\frac12\bigl(U_{h,1/2}^- + U_{h,1/2}^+\bigr)
- \frac{1}{2\alpha}\bigl(F(U_{h,1/2}^+)-F(U_{h,1/2}^-)\bigr), \\
\Phi^{(k)}(U_I^{(k)})
=
\begin{pmatrix}
q_I^{(k)} - (\rho_I^{(k)})^{\gamma+1} - \hat\delta_{j,k} \\
q_I^{(k)} - \rho_I^{(k)}
\end{pmatrix}.
\end{gather}

The density flux $\hat\delta_{j,k}$ is determined by solving the following optimization problem:
\begin{equation}
\label{eq:maximization_of_velocity}
\left\{
\begin{array}{c}
\text{maximize } 
\displaystyle{\sum_{e_k \in \mathrm{In}_j} \hat\delta_{j,k}}
\quad \text{subject to} \\
\hat\delta_{j,k} \in [0,\delta_{j,k}^{\textup{Max}}], \qquad
\displaystyle{\hat\delta_{j,n} = \sum_{e_k\in \mathrm{In}_j} a_{n,k}\,\hat\delta_{j,k}},
\quad e_n \in \mathrm{Out}_j,
\end{array}
\right.
\end{equation}
where $\delta_{j,k}^{\textup{Max}}$ denotes the maximum admissible density flux (see \ref{sec:coupling} for the definition) and $A_j=(a_{n,k})$ is the distribution matrix at the junction $\mathrm{P}_j$.

In the above construction, the coupling conditions~(C1)--(C5) are incorporated as follows.
Condition~(C1) is expressed by the conservation constraint conditions in Eq.~\eqref{eq:maximization_of_velocity}.
The minimization of $\|D_{\textup{in}}^{(k)}\|_{\mathbb{R}^2}$ ensures that the waves generated at the junction satisfy~(C2).
Condition~(C3) is also included in Eq.~\eqref{eq:maximization_of_velocity}.
The relation
$q_I^{(k)} - (\rho_I^{(k)})^{\gamma+1} = \hat\delta_{j,k}$ together with
Eq.~\eqref{eq:maximization_of_velocity} enforces the maximization of the total density flux on incoming roads~(C4).
Finally, the condition $q_I^{(k)} = \rho_I^{(k)}$ ensures the maximization of the velocity on outgoing roads~(C5).

\subsection{Computational algorithm}

\rev{
The numerical procedure takes the network configuration and the prescribed initial and boundary conditions as input, and advances the ARZ dynamics on the network through the junction coupling and DG--SSPRK discretization. The resulting time-dependent density field serves as the basis for the congestion-cluster extraction and statistical analysis presented in Section~\ref{sec:results}.
}

The procedure for computing the solution at the next time level $U_h^{i+1}$ from $U_h^{i}$ on each road $e_k$ is summarized as follows.
\begin{enumerate}[(a)]
    \item Compute the maximum admissible density flux $\delta_{j,k}^{\textup{Max}}$ using $U_h^{i}$.
    \item Determine the boundary conditions by solving the corresponding minimization problem at each junction.
    \item Solve the DG scheme~\eqref{eq:dg_scheme} to compute $L(U_h^{i})$.
    \item Apply the first stage of the SSPRK3 method~\eqref{eq:ssprk1} to obtain $U_h^{(1)}$.
    \item Compute $U_h^{i+1}$ by repeating steps (a)--(d) in accordance with the remaining stages of the SSPRK3 method, namely Eqs.~\eqref{eq:ssprk2} and \eqref{eq:ssprk3}.
\end{enumerate}

%
\section{Numerical results}
\label{sec:results}

In this section, we present numerical results obtained by solving the ARZ continuum traffic flow model on a large-scale road network.
Our primary objective is to examine whether the scale-free statistical properties of traffic congestion clusters, reported in the empirical study by Zhang \textit{et al.}~\cite{zhang2019}, can be reproduced within a continuum modeling framework.
In particular, we analyze the spatiotemporal structure of congestion clusters and the frequency distribution of their size $S$.
The methodology of extracting congestion clusters follows the spatiotemporal connectivity concept illustrated in Fig.~\ref{Fig:intro}.
\rev{
Particular attention is paid to the resulting power-law distributions and their finite-size scaling behavior. To clarify the organization of the evaluation, we first describe the network setting and boundary driving, then present representative spatiotemporal congestion patterns, and finally examine the cluster-size statistics together with their dependence on network size.
}

\subsection{Road network and boundary conditions}
\label{subsec:network}

\begin{figure}[tb]
    \centering 
    \includegraphics[clip, scale=1.2]{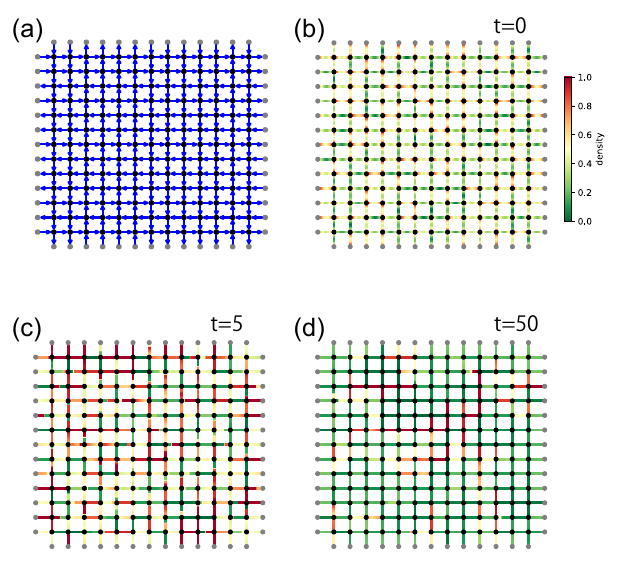} 
    \caption{Visualization of density distributions on a lattice network. (a) Overview of the lattice network used in the simulation. Panels (b), (c), and (d) show the density $\rho$ on each link at times $t=0$, $t=5$, and $t=50$, respectively, illustrating the generation, propagation, and dissipation of high-density regions (the color bar indicates density).}
    \label{Fig:density}
\end{figure}

The numerical simulations are performed on a lattice-shaped road network shown in Fig.~\ref{Fig:density}(a).
The presented network consists of $13\times13=169$ intersections (nodes), each connected by road links to its nearest neighbors. We denote by $N$ the total number of intersections (nodes) in the network, i.e., $N=169$ in the case of Fig.~\ref{Fig:density}(a).

All links are assumed to be one-way roads, and their directions are assigned randomly at the beginning of each simulation.
Although the geometric structure is regular, the random assignment of directions introduces significant asymmetry and heterogeneity in traffic flow.
In particular, it creates local imbalances between incoming and outgoing connectivity, which is a key ingredient for producing nonuniform density patterns under junction coupling.

\rev{
Although the network geometry is lattice-like and may resemble an urban road grid, the present setting is intended as a simplified directed network model rather than a direct model of real urban traffic operation. In particular, all links are treated as one-way roads, and neither traffic signals nor bidirectional intersection operations are explicitly incorporated. Therefore the present setup should be interpreted as a controlled continuum-model system designed to isolate the effect of networked macroscopic flow dynamics on congestion-cluster statistics.
}

At the outer boundary of the network, we impose time-periodic Dirichlet boundary conditions.
More specifically, the boundary density at each outer node is prescribed as a periodic piecewise-defined time-dependent signal characterized by a baseline value, a driving period, and an oscillation amplitude.
The initial value of a boundary node is set to the initial value of the road for which the node serves as either the source or the target, as described later.
After that, at fixed time intervals, the boundary condition repeatedly cycles through holding constant, decreasing to the target value, holding constant, and increasing to the target value. 
The target value during the decreasing and increasing intervals is the same for all nodes. 
The corresponding boundary state is chosen consistently with the ARZ model so as to remain within the admissible invariant region.
These boundary conditions allow inflow and outflow of vehicles and thus render the system non-conservative with respect to the total number of vehicles inside the network.
As a consequence, the traffic state does not converge to a static steady state; instead, it remains in a driven, non-equilibrium regime in which congestion can repeatedly emerge and dissipate.
This setting provides a controlled numerical analogue of time-varying demand in urban traffic, where boundary supply and sink effects are unavoidable.
From the viewpoint of congestion statistics, the non-stationary driving is essential: it prevents the dynamics from becoming trapped in a trivial uniform state and yields a sufficiently rich ensemble of congestion events for reliable statistical characterization of $F(S)$.

Initial conditions are generated randomly, as illustrated in Fig.~\ref{Fig:density}(b).
The density on each link is assigned independently with random fluctuations around a prescribed mean value, which is set to $\rho=0.5$ in the present study.
By avoiding artificially imposed large-scale structures at $t=0$, we ensure that the observed congestion patterns are produced by the intrinsic dynamics of the ARZ model on a coupled network.
In all simulations, the computed density remains within the admissible invariant region discussed in \ref{sec:coupling}, and the junction coupling rules maintain physically consistent propagation directions of waves.

\subsection{Extracting congestion clusters}
\label{subsec:clusterdef}

To characterize traffic congestion from a resilience perspective, we adopt the spatiotemporal congestion-cluster framework proposed by Zhang \textit{et al.}~\cite{zhang2019}.
In this approach, congestion is treated not as a purely instantaneous spatial pattern but as a connected object in space and time, as schematically illustrated in Fig.~\ref{Fig:intro}.
This viewpoint enables us to treat congestion as a connected component in space and time, rather than a purely instantaneous spatial pattern.

For a given density threshold $\rho_{\mathrm{thres}}$, we first identify congested links at each time step and then construct spatiotemporal clusters by combining spatial and temporal connectivity.
The cluster extraction procedure is defined as follows:
\begin{enumerate}[(1)]
    \item At each time instant, road links whose average density exceeds $\rho_{\mathrm{thres}}$ are labeled as congested links.
    \item Congested links that are spatially adjacent through network nodes at the same time are grouped into spatial clusters.
    \item Temporal connectivity is introduced by linking congested links (or spatial clusters) at time $t$ to those at time $t+\Delta t$ whenever they share at least one common node and the node has nonzero degree within the congested subgraph.
    \item By combining spatial and temporal adjacency, we obtain a spatiotemporal graph; each connected component of this graph is defined as a congestion cluster.
    \item The cluster size $S$ is defined as the total number of congested links contained in the cluster (summed over all time slices within the component).
\end{enumerate}
This definition is consistent with the $d+1$ dimensional cluster concept of \cite{zhang2019}.
In practice, the choice of $\rho_{\mathrm{thres}}$ affects the set of congested links.
If $\rho_{\mathrm{thres}}$ is chosen too small, a single large congested component tends to span the network over the entire observation period, effectively producing one dominant cluster that masks the statistical diversity of events. Since too small values of $\rho_{\mathrm{thres}}$ tend to produce a single system-spanning cluster that persists over the entire observation period, thereby obscuring meaningful statistical characterization, we restrict attention to $\rho_{\mathrm{thres}} \ge 0.9$. In this regime, multiple clusters of different scales coexist, and no fine tuning of $\rho_{\mathrm{thres}}$ is required to observe power-law behavior. We confirm that the resulting scaling exponents remain robust under moderate variations of the threshold.

\subsection{Evolution of density fields}
\label{subsec:densityevo}

Figures~\ref{Fig:density}(b)--(d) show snapshots of the density field at times $t=0$, $t=5$, and $t=50$, respectively.
Starting from random initial conditions , localized high-density regions emerge spontaneously.
This behavior can be understood as a network-level consequence of nonlinear wave propagation in the ARZ model combined with junction coupling:
waves generated at one location are redistributed at junctions, which can amplify local density when multiple inflows compete for limited outgoing capacity, and can also relieve congestion when outflow becomes available.
As time evolves, congested regions propagate along directed links, interact with other waves, and undergo merging and splitting events.

Because time-dependent Dirichlet boundary conditions are imposed at the outer nodes, the system is continuously driven.
As a result, congestion events do not simply decay and vanish permanently; instead, new congestion is repeatedly generated.
This sustained non-equilibrium regime is particularly important for observing a wide range of congestion scales: small, short-lived events coexist with rare but persistent large events.
From the numerical perspective, the repeated excitation also reduces sensitivity to the particular random initial condition, enabling the aggregation of statistics across multiple runs.

\subsection{Spatiotemporal congestion clusters}
\label{subsec:clustershape}

\begin{figure}[p]
    \centering 
    \includegraphics[clip, scale=1]{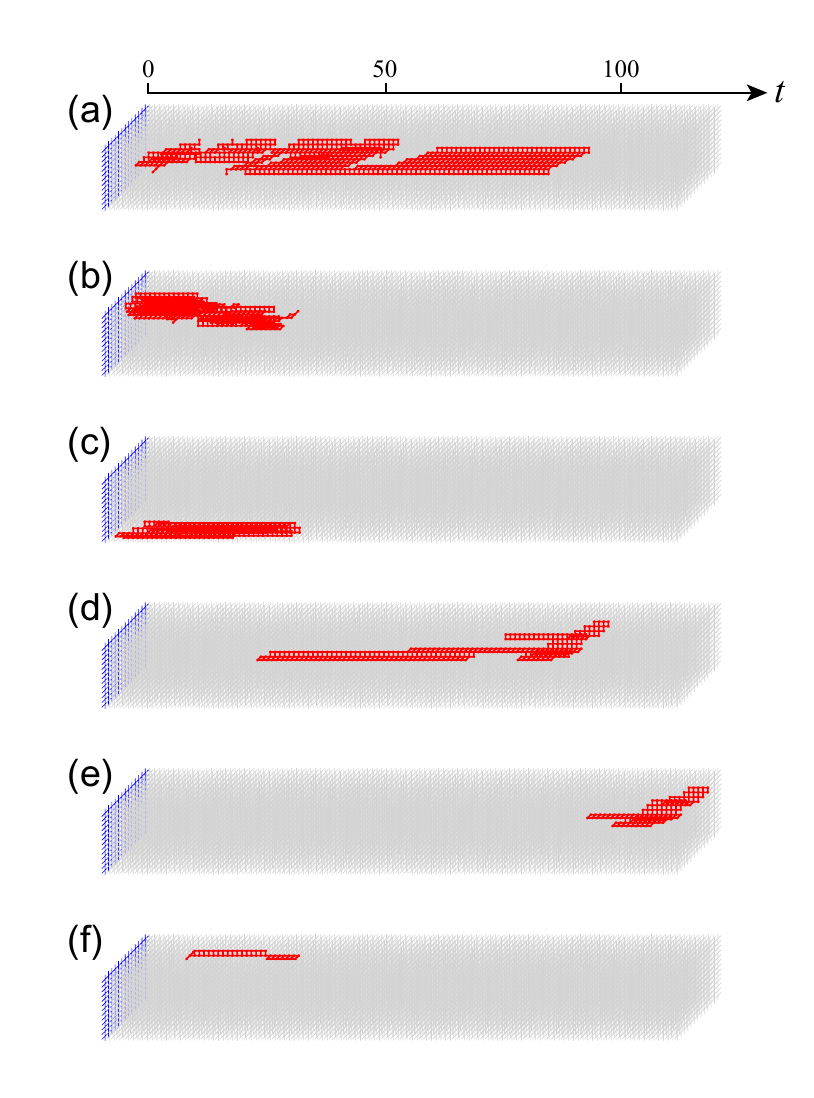} 
    \caption{Examples of spatiotemporal structures of extracted congestion clusters. Links whose density exceeds the threshold $\rho_{\mathrm{thres}} = 0.9$ are classified as congested, and clusters are defined based on spatial and temporal connectivity. The lattice at $t=0$ is shown in blue, and the horizontal axis represents time $t$. Panels (a)-(f) display clusters ordered by size, corresponding to the 1st, 2nd, 3rd, 5th, 10th, and 20th largest clusters, respectively.}
    \label{Fig:clusters}
\end{figure}

Examples of extracted spatiotemporal congestion clusters for $\rho_{\mathrm{thres}}=0.9$ are shown in Fig.~\ref{Fig:clusters}.
Panels (a)--(f) correspond to clusters ranked by size, specifically the 1st, 2nd, 3rd, 5th, 10th, and 20th largest clusters.
The horizontal direction represents time, and each panel visualizes how a cluster persists and evolves across time slices.
The fact that clusters form connected objects in space and time highlights that congestion is not a purely local phenomenon, but one that can propagate through the network and maintain temporal coherence.

Large clusters exhibit complex spatiotemporal behavior.
Rather than remaining localized at a fixed position, they expand and contract in space, and often undergo merging and splitting processes.
Such events can be interpreted as congestion waves interacting at junctions: local bottlenecks create high-density patches, while alternative outgoing links can redirect flow and fragment a congested region.
The coexistence of diverse cluster morphologies across scales provides a qualitative basis for the scale-free statistics discussed in the next subsection.

\subsection{Cluster statistics}
\label{subsec:clusterstats}

To obtain statistically robust results, simulations are carried out for the time interval $t=0$ to $t=120$, repeating the calculations 60 times with different random assignments of link directions and initial conditions. All congestion clusters extracted from these runs are aggregated to compute the frequency distribution of cluster size $S$. 
We denote by $F(S)$ the frequency distribution of cluster sizes. 
\rev{
To examine the size dependence more systematically, we compare regular lattice networks with $N=9$, $25$, $64$ and $169$. In addition, to assess the effect of topology, we also include an irregular network with $N=64$, obtained by randomly removing 10\% of the edges from the original lattice.}

\begin{figure}[t]
    \centering 
    \includegraphics[clip, scale=1.2]{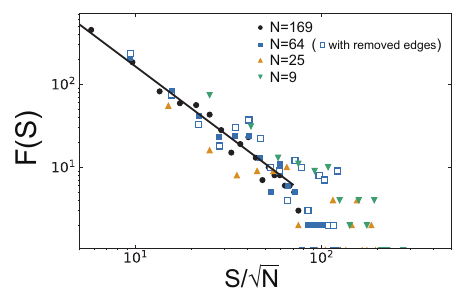} 
    \caption{Frequency distribution of congestion cluster size for networks of different sizes. 
Distribution $F(S)$ of the spatiotemporal cluster size $S$ plotted against the scaled variable $S/\sqrt{N}$. Here, $N$ denotes the total number of intersections in the network. 
\rev{ Results are shown for $N=9$, $25$, $64$ and $169$ on regular lattice networks, together with an irregular network of $N=64$ with removing 10\% of the edges (shown with open square).} Straight lines indicate linear fits in the log--log plane, performed over the range where the number of observed clusters remains statistically significant.
}
    \label{Fig:hist}
\end{figure}

Figure~\ref{Fig:hist} presents the frequency distribution of cluster size $F(S)$ for networks of different sizes.
The linear fits in the log--log plot are performed over the scaling range where each bin contains a sufficient number of clusters to ensure statistical reliability, excluding the large-size tail influenced by finite-size cutoff.
Here, $N$ denotes the total number of intersections in the network.
For a lattice of size $L\times L$, we have $N=L^2$, so that $\sqrt{N}$ represents the linear system size.
Instead of plotting the raw distribution, we rescale the horizontal axis by $\sqrt{N}$ and display $F(S)$ as a function of $S/\sqrt{N}$.
Remarkably, the data for different network sizes collapse approximately onto a single straight line in log--log coordinates.
The resulting slope is approximately $\alpha \approx 1.85$ for $F(S)$.
This collapse suggests a finite-size scaling form,
\begin{equation}
F(S;N) = S^{-\alpha} f\!\left(\frac{S}{S_c(N)}\right),
\end{equation}
with a cutoff scale satisfying
\begin{equation}
S_c(N) \propto \sqrt{N}.
\end{equation}

The emergence of a scale factor proportional to $\sqrt{N}$ admits a natural geometric interpretation. 
Since the underlying network is a two-dimensional lattice with $N = L^{2}$ nodes, the only intrinsic macroscopic length scale of the system is the linear size $L \sim \sqrt{N}$. 
The observed data collapse therefore indicates that the characteristic scale of large congestion clusters is governed by this linear system size rather than by any finite intrinsic correlation length. 
In other words, the continuum network dynamics do not exhibit a characteristic congestion scale smaller than the network size itself, and the cutoff in the cluster-size distribution is determined primarily by the global system size.

We emphasize that the cluster size $S$ analyzed here is defined as a spatiotemporal quantity, namely the total number of congested links accumulated over the entire lifetime of a cluster.
Thus, $S$ corresponds to a $(d+1)$-dimensional connected component rather than a purely spatial one.

In the language of critical phenomena, this behavior indicates that the effective correlation scale of congestion becomes comparable to the system size.
Although we do not directly measure the spatial correlation length, the observed finite-size collapse suggests that congestion clusters are not governed by a fixed intrinsic scale but instead expand up to a scale limited only by the network size.

We further note that the estimated exponents $\alpha$ remain within $-1.69\sim -1.84$ when varying the density threshold $\rho_{\mathrm{thres}}$ from $0.9$ to $0.94$.
Therefore, the scale-free behavior and the finite-size scaling collapse are robust with respect to the threshold choice.

\rev{
To assess whether the observed scale-free behavior is specific to the regular lattice topology, we also performed supplementary simulations on an irregular network with $N=64$, constructed by randomly removing 10\% of the edges from the original lattice. We confirmed that, at least within the range examined here, the cluster-size distribution still exhibits a power-law decay, with an estimated exponent close to that in the regular-lattice case. As also seen in Fig.~6, the $N=64$ data with removed edges are broadly aligned with the same decaying trend as the regular-lattice cases. This suggests that the emergence of scale-free-like congestion clusters is not merely an artifact of the perfectly regular lattice topology.
}

Notably, the estimated exponents are quantitatively in good agreement with the 
highway results reported in \cite{zhang2019}, suggesting that the 
scale-free structure observed here is consistent with corridor-like 
traffic dynamics. 
\rev{
The present directed network configuration, lacking explicit signal control, bidirectional street operation, and other heterogeneous urban constraints, naturally captures features closer to highway systems than to fully developed urban traffic networks.
We note, however, that the quantitative values of the exponents may also depend on the adopted junction coupling rule, though this dependence is left for future study.
}

Although we refrain from claiming strict self-organized criticality, 
the observed coexistence of scale-free and system-size--dependent distributions  
strongly resembles critical-like behavior in driven non-equilibrium systems. 
In the present continuum framework, such scale-free dynamics emerge 
without explicit parameter tuning, suggesting that scale invariance arises 
as an intrinsic property of nonlinear networked traffic flow.

\section{Concluding remarks}
\label{sec:clusterstats}
In this study, we examined the spatiotemporal statistics of traffic congestion on large-scale networks within a macroscopic continuum framework. Solving the Aw-Rascle-Zhang model on lattice networks of varying sizes, we extracted congestion clusters and analyzed the distribution of their size $S$. The cluster size exhibits approximate power-law behavior over a wide range.

More importantly, when rescaled by the linear system size $L\sim\sqrt{N}$, the distributions for different network sizes collapse onto a nearly universal curve, indicating finite-size scaling proportional to $L$. The scaling behavior remains robust under variations of the density threshold. These results demonstrate that scale-free spatiotemporal congestion statistics and their system-size dependence can emerge intrinsically from networked continuum traffic dynamics under non-equilibrium conditions.

\rev{ 
Motivated by the scale-free congestion statistics identified in real urban traffic data~\cite{zhang2019}, the present study examines whether this key statistical property can be reproduced by a continuum model of macroscopic traffic flow on networks. While our aim is not direct quantitative comparison to a specific real dataset, the resulting scaling exponents can still be compared with the empirical values reported for real traffic networks. Although the exponent obtained here differs from the value reported for urban traffic, it is broadly consistent with the exponent reported for highway traffic in~\cite{zhang2019}. The remaining quantitative difference from the urban case is likely attributable to the simplified network setting. In particular, the present model does not explicitly incorporate traffic signals, heterogeneous road capacities, route choice, or data-calibrated demand fluctuations, all of which may affect the quantitative values of the scaling exponents.
At the same time, supplementary simulations on an irregular network suggest that the scale-free-like behavior and the corresponding exponent are not strongly altered by a moderate deviation from the regular lattice topology. A systematic clarification of how the scaling structure depends on network topology, boundary driving, and junction coupling rules remains an important direction for future work. 

An important extension is to introduce more realistic urban operating elements, such as traffic signals and bidirectional intersection structures, and to examine how these features modify the scaling exponents of congestion clusters.
From the continuum-modeling viewpoint, such extensions require development of the formulation of more realistic coupling rules at junctions. Constructing and analyzing such coupling conditions remains a challenging problem in continuum traffic modeling on networks.
}
Extending the present analysis to heterogeneous or partially disrupted networks may further illuminate the relation between congestion scaling and complex-network structure, in connection with percolation and robustness theories developed for networked systems \cite{cohen2010, li2015}.

\rev{
Another important extension is to examine mixed traffic systems involving connected and autonomous vehicles (CAVs), semi-autonomous vehicles (SAVs), and human-driven vehicles (HDVs). Such vehicles may modify lane usage, headway, lane-changing, and junction behavior in realistic traffic environments \cite{Dubey2025}, which in turn may affect the formation and evolution of spatiotemporal congestion clusters. In a continuum-modeling framework, these effects could be represented through modified constitutive relations, anticipation terms, junction coupling rules, or additional control terms \cite{Imran2024,Mohammadian2023}. Clarifying how these factors influence the statistical properties of congestion clusters, including the power-law behavior observed in the present study, remains an important topic for future research.
}

Overall, the present findings indicate that scale-free spatiotemporal congestion dynamics can arise within a macroscopic continuum description of networked traffic flow. 
The observed power-law statistics and finite-size scaling suggest that large-scale statistical features of urban congestion do not necessarily require detailed microscopic modeling, but can emerge from coarse-grained network dynamics.

\appendix

\section{Mathematical remarks on the ARZ model}
\label{sec:math}

We briefly review the mathematical results for the ARZ model \eqref{Eq:AR_model}.  
Throughout this appendix, no particular functional form relating the pressure to the density,
such as Eq.~\eqref{eq:p1}, is assumed. 
Thus, $p=p(\rho)$ is assumed only to be smooth, increasing function of $\rho$. 

Recall that the ARZ model is described as the conservative form 
Eqs.~\eqref{eq:conservation_of_laws0} and \eqref{eq:conservation_of_laws}. The Jacobian matrix $DF(z)$ of $F(z)$ at $z=(z_1,z_2)^\top$ is given by
\begin{equation}
    DF(z) = \begin{pmatrix}
        -p(z_1)-z_1p'(z_1) & 1 \\[1mm]
        -\left(\dfrac{z_2}{z_1}\right)^2 - z_2p'(z_1) & 2\dfrac{z_2}{z_1}-p(z_1)
    \end{pmatrix},
\end{equation}
and its eigenvalues are
\begin{equation}
    \lambda_1(z) = \frac{z_2}{z_1} - p(z_1) - z_1p'(z_1), \qquad
    \lambda_2(z) = \frac{z_2}{z_1} - p(z_1).
\end{equation}
In physical notation, using Eq.~\eqref{eq:velocity}, we have
\begin{equation}
    \lambda_1(U) = v - \rho p'(\rho) < v = \lambda_2(U).
\end{equation}
Therefore, the ARZ model is strictly hyperbolic in the sense of Lax. 
Moreover, every wave propagation speed is at most equal to $v$, which is a desirable property for a traffic flow model.

We next recall further properties of the ARZ model as a hyperbolic system, as reported in \cite{AwRascle2000}; see, for example, \cite[Section~11]{Evans2010}, \cite{Dafermos2016}, and \cite{HoldenRisebro2015} for the basic theory of hyperbolic systems of conservation laws.
Since the corresponding (right) eigenvectors associated with $\lambda_1(z)$ and
$\lambda_2(z)$ are given by
\begin{equation}
    r_1(z) = \begin{pmatrix}
        1 \\
        \dfrac{z_2}{z_1}
    \end{pmatrix}, \qquad
    r_2(z) = \begin{pmatrix}
        1 \\
        \dfrac{z_2}{z_1} + z_1 p'(z_1)
    \end{pmatrix},
\end{equation}
we compute
\begin{equation}
    \nabla \lambda_1(z)\cdot r_1(z)
    = -\bigl(2p'(z_1) + z_1 p''(z_1)\bigr),
    \qquad
    \nabla \lambda_2(z)\cdot r_2(z) = 0.
\end{equation}
Therefore, the pair $(\lambda_2,r_2)$ is always linearly degenerate.
Henceforth, we additionally assume that
\begin{equation}
    \label{Eq:strictly_convex}
    2p'(\rho) + \rho p''(\rho) \neq 0
    \quad \mbox{for all } \rho > 0,
\end{equation}
which means that the function $\rho \mapsto \rho p(\rho)$ is strictly convex.
Under this assumption, the pair $(\lambda_1,r_1)$ is genuinely nonlinear,
and therefore shocks or rarefaction waves may occur.

Moreover, \cite{AwRascle2000} observed that the Riemann invariants (in the sense of Lax) are given by
\begin{equation}
    w_1 = v + p(\rho), \qquad w_2 = v.
\end{equation}
Recall that a Riemann invariant $w_k$ is constant along integral curves of the $k$-th family. This implies that the $1$-rarefaction curve in the $(\rho,q)$-plane
is a straight line of the form $q = \rho\bigl(v + p(\rho)\bigr)$,
whose slope $v + p(\rho)$ is constant. The curve of the second family in the
$(\rho,q)$-plane is also described by $q = \rho\bigl(v + p(\rho)\bigr)$, 
where $v$ is interpreted as a constant. Therefore, the curve of the second family through a given
point $(\rho_0,q_0)$ is given by
\begin{equation}
    \label{Eq:curve_of_second_family}
    q = \rho\left(\frac{q_0}{\rho_0} - p(\rho_0) + p(\rho)\right).
\end{equation}

Next, we consider invariant regions. Recall that $\rho_{\mathrm{max}} > 0$ is a prescribed maximal
density and $v_{\mathrm{max}} > 0$ a prescribed maximal velocity. As noted in
\cite{AwRascle2000}, if
\begin{equation}
    \label{Eq:pressure_proportional_to_exponential}
    p(\rho) \propto \rho^{\gamma}
    \quad \text{as } \rho \to 0
\end{equation}
for some $\gamma > 0$, then the region
\begin{equation}
    \mathcal{R}
    = \bigl\{(\rho,v) \mid \rho \ge 0,\ 0 \le v \le v_{\mathrm{max}} - p(\rho)\bigr\}
\end{equation}
is invariant for the Riemann problem. Here we assume that
$v_{\mathrm{max}} - p(\rho_{\mathrm{max}}) = 0$.
Since $p(\rho)$ is assumed to be an increasing function, we have
$v_{\mathrm{max}} \ge p(\rho)$ for all $0 \le \rho \le \rho_{\mathrm{max}}$.
Therefore,
\[
    \rho\,p(\rho) \le q = \rho\bigl(v + p(\rho)\bigr)
    \le \rho\,v_{\mathrm{max}}
    \quad \text{for all } (\rho,v) \in \mathcal{R}.
\]
This implies that
\begin{equation}
    \bigl\{(\rho,q) \mid \rho\,p(\rho) \le q \le \rho\,p(\rho_{\mathrm{max}})\bigr\}
\end{equation}
is an invariant region for the ARZ model \eqref{Eq:AR_model}.

\section{Additional remarks on coupling conditions}
\label{sec:coupling}

In this appendix, we precisely describe what is meant by the coupling conditions (C4) and (C5)
introduced in Subsection~2.3. In fact, (C4) and (C5) should be interpreted as (C4$'$) and (C5$'$)
defined below. 

We consider an $N$--$M$ junction, and assume that coupling conditions (C1), (C2) and (C3) are satisfied. 
Recall that we are assuming Eq.~\eqref{eq:p1} relating the pressure to the density. Then, Eqs.~\eqref{Eq:strictly_convex} and \eqref{Eq:pressure_proportional_to_exponential} are satisfied. Moreover, the invariant region is given by 
\begin{equation}
\mathcal{D}=
    \bigl\{(\rho,q) \mid 
    \rho^{1+\gamma}\le q\le \rho
    \bigr\}.
\end{equation}

In \cite{GaravelloPiccoli2006}, the authors focus on the density flux (the first component of the
flux $F(U)$) and characterize the admissible final states of the density flux for the Riemann
problem on incoming and outgoing roads.
Following \cite{GaravelloPiccoli2006},  we define two subdomains of $\mathcal{D}$ as
\begin{align}
\mathcal{D}_1 &= \{(\rho, q)\in\mathcal{D} \mid  (\gamma+1)\rho^{\gamma+1}\le q\le \rho\}, \\ 
\mathcal{D}_2 &= \{(\rho, q)\in\mathcal{D} \mid  \rho^{\gamma+1}\le q\le (\gamma+1)\rho^{\gamma+1}\},   
\end{align}
and consider the Riemann problem with the initial state $(\rho_0^{(n)}, q_0^{(n)})\in\mathcal{D}$ at the $n$-th incoming road $e_n$ and $(\rho_0^{(m)}, q_0^{(m)})\in\mathcal{D}$ at the $m$-th outgoing road $e_m$, where we write  $\rho_0^{(n)}=\rho^{(n)}(x,0)$ for example.

The density fluxes in the $n$-th incoming road and the $m$-th outgoing road are denoted by
$\delta_n$ and $\delta_m$, respectively. That is, for example, we set
\begin{equation}
    \delta_n = q^{(n)} - \bigl(\rho^{(n)}\bigr)^{\gamma+1}.
\end{equation}

It is proved that $\delta_n \in [0, \delta_n^{\text{Max}}]$ for incoming roads \cite[Proposition 5.1]{GaravelloPiccoli2006}, where 
\begin{equation}
    \delta_n^{\text{Max}} = \left\{\begin{array}{cl} 
        q^{(n)}_0-(\rho^{(n)}_0)^{\gamma+1} & \text{if }(\rho^{(n)}_0, q^{(n)}_0) \in \mathcal{D}_1, \\ 
        \gamma\left(\dfrac{q^{(n)}_0}{(\gamma+1)\rho^{(n)}_0}\right)^{\frac{\gamma+1}{\gamma}} & \text{if }(\rho^{(n)}_0, q^{(n)}_0) \in \mathcal{D}_2 .
    \end{array}\right.
\end{equation}

Similarly, for the $m$-th outgoing road, the density flux satisfies $\delta_m \in [0,\delta_m^{\mathrm{Max}}]$,  
see \cite[Proposition~5.1]{GaravelloPiccoli2006}, where $\delta_m^{\mathrm{Max}}$ is defined as follows.
If the curve of the second family through $(\rho_0^{(m)}, q_0^{(m)})$ is entirely contained in
$\mathcal{D}_1$, then
\begin{equation}
    \label{Eq:maximum_flux_of_density_in_outgoing_road1}
    \delta_m^{\mathrm{Max}}
    =
    \frac{\gamma}{1+\gamma}
    \left(1-\frac{\gamma}{\gamma+1}\right)^{\frac{1}{\gamma}}.
\end{equation}
Otherwise, $\delta_m^{\mathrm{Max}}$ is given by
\begin{equation}
    \label{Eq:maximum_flux_of_density_in_outgoing_road2}
    \delta_m^{\mathrm{Max}}
    =
    \frac{q_0^{(m)}-(\rho_0^{(m)})^{\gamma+1}}{\rho_0^{(m)}}
    \left(
        1-\frac{q_0^{(m)}-(\rho_0^{(m)})^{\gamma+1}}{\rho_0^{(m)}}
    \right)^{\frac{1}{\gamma}}.
\end{equation}

From the admissible state of density flux, it has been proposed to determine the solution to the Riemann problem on incoming roads uniquely by the following condition:
\begin{itemize} 
    \item [(C4$'$)] The sum of flux of density in incoming roads $\displaystyle\sum_{n=1}^N \delta_n$ at the junction must be maximized. 
\end{itemize}

\begin{figure}[tb]
    \centering 
    \includegraphics[clip, scale=1.5]{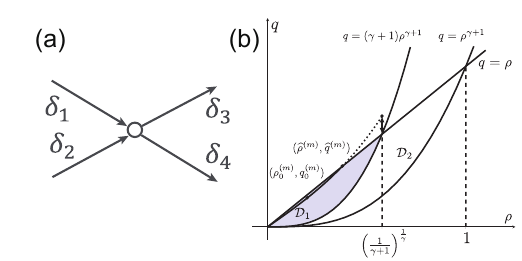} 
    \caption{(a) Conceptual diagram of the density flux $\delta_i$ at a junction. (b) Invariant domain $\mathcal{D}$ and its subdomains $\mathcal{D}_1$ and $\mathcal{D}_2$ in the $(\rho,q)$ plane, together with the second-family curve passing through a given state $(\rho^{(m)}_0,q^{(m)}_0)$. The straight line $q=\rho$ corresponds to the velocity-maximization condition on outgoing roads (C5).}
    \label{Fig:model}
\end{figure}

Moreover, it has been proposed to introduce one of three different additional rules in order
to determine a unique solution on outgoing roads. One of the proposed conditions is to maximize the velocity on each outgoing road \cite[Proposition~5.4]{GaravelloPiccoli2006}. That is, 
\begin{itemize}
    \item[(C5$'$)] 
    The final state $(\widehat{q}^{(m)}, \widehat{\rho}^{(m)})$ on the $m$-th
    outgoing road satisfies $\widehat{q}^{(m)} = \widehat{\rho}^{(m)}$
\end{itemize}
Under this assumption, the quantity $\delta_m^{\text{Max}}$ can be determined by following the curve of the second family issuing from $(\rho_0^{(m)}, q_0^{(m)}) \in \mathcal{D}$ until it intersects
the straight line $q = \rho$ (see Figure~\ref{Fig:model}).

Recall that the curve of the second family is given by Eq.~\eqref{Eq:curve_of_second_family}.
At this stage, we focus on this curve in the domain 
\[
    0\le \rho \le \left(\dfrac{1}{\gamma+1}\right)^{\frac{1}{\gamma}}.
\]
Then, we see that 
\begin{align}
            \rho\left(\dfrac{q_0^{(m)}}{\rho_0^{(m)}} - p(\rho_0^{(m)}) + p(\rho)\right) - (\gamma+1)\rho^{\gamma+1} &= \rho\left(\dfrac{q_0^{(m)}}{\rho_0^{(m)}}-(\rho_0^{(m)})^{\gamma} - \gamma\rho^\gamma\right) \\ 
        &\ge \rho\left(\dfrac{q_0^{(m)}-(\rho_0^{(m)})^{\gamma+1}}{\rho_0^{(m)}} - \dfrac{\gamma}{\gamma+1}\right)
    \end{align}

Therefore, we infer that the curve of the second family through $(\rho_0^{(m)}, q_0^{(m)})$ is completely inside $\mathcal{D}_1$ if and only if 
\[
    \dfrac{q_0^{(m)}-(\rho_0^{(m)})^{\gamma+1}}{\rho_0^{(m)}} - \dfrac{\gamma}{\gamma+1} > 0.
\]
Consequently, 
Eqs.~\eqref{Eq:maximum_flux_of_density_in_outgoing_road1} and \eqref{Eq:maximum_flux_of_density_in_outgoing_road2} are  summarized as 
\begin{equation}
    \delta_m^{\text{Max}} = \left\{\begin{array}{cc} 
        \dfrac{\gamma}{1+\gamma}\left(1-\dfrac{\gamma}{\gamma+1}\right)^{\frac{1}{\gamma}}, & \text{if}\quad \dfrac{q_0^{(m)}-(\rho_0^{(m)})^{\gamma+1}}{\rho_0^{(m)}} > \dfrac{\gamma}{\gamma+1}, \\ 
        \dfrac{q^{(m)}_0-(\rho^{(m)}_0)^{\gamma+1}}{\rho^{(m)}_0}\left(1-\dfrac{q^{(m)}_0 - (\rho^{(m)}_0)^{\gamma+1}}{\rho^{(m)}_0}\right)^{\frac{1}{\gamma}}, & \text{otherwise}.
    \end{array}\right.
\end{equation}

In the optimization problem 
\eqref{eq:maximization_of_velocity} associated with the node
$\mathrm{P}_j$, the quantity $\delta_{j,k}^{\mathrm{Max}}$ is chosen as either
$\delta_k^{\mathrm{Max}}$ or $\delta_m^{\mathrm{Max}}$ defined above, depending on whether the
road $e_k$ under consideration is incoming or outgoing.

\section*{Acknowledgements}

This work was partially supported by JSPS KAKENHI Grant Number 21H04431 (Grant-in-Aid for Scientific Research (A)). 

\section*{Author Contributions (CRediT)}

Yuki Chiba: Methodology, Software, Investigation,
Data curation, Formal analysis, Writing - review \& editing.

Yuki Ueda: Formal analysis, Methodology,
Software, Investigation,
Writing - review \& editing.

Norikazu Saito: Conceptualization, Supervision,
Project administration,
Writing - review \& editing.

Hiroaki Yoshida: Conceptualization, Formal analysis,
Project administration, Supervision,
Writing - original draft, Writing - review \& editing.

\section*{Data Availability}
The simulation data and analysis code supporting this study are available from the corresponding author upon reasonable request.

\section*{Declaration of Interests}
The authors declare that they have no known competing financial interests or personal relationships that could have appeared to influence the work reported in this paper.

\section*{Declaration of Generative AI and AI-assisted Technologies in the Manuscript Preparation Process}
During the preparation of this manuscript, the authors used generative AI tools to assist with English language refinement and clarity. 
All scientific content, analysis, and conclusions were developed and verified by the authors, who take full responsibility for the integrity of the manuscript.


\end{document}